\documentclass[prl,twocolumn,preprintnumbers,amsmath,amssymb,superscriptaddress]{revtex4}

\usepackage{graphicx}
\usepackage{dcolumn}
\usepackage{bm}
\usepackage[usenames]{color}

\begin{document}

\title{Pseudomagnetic fields and ballistic transport in a
suspended graphene sheet}

\author{M. M. Fogler}
\affiliation{Department of Physics, University of California
San Diego, La Jolla, 9500 Gilman Dr. CA 92093, USA}

\author{F. Guinea}
\affiliation{Instituto de Ciencia de Materiales de Madrid
(CSIC), Sor Juana In\'es de la Cruz 3, Madrid 28049, Spain and
Donostia International Physics Center (DIPC), Paseo Manuel de
Lendiz\'abal 4, San Sebasti\'an, E-20018, Spain}

\author{M. I. Katsnelson}
\affiliation{Institute for Molecules and Materials, Radboud
University Nijmegen, Heijendaalseweg 135, 6525 AJ, Nijmegen, The
Netherlands}

\begin{abstract}

We study a suspended graphene sheet subject to the electric field of
a gate underneath. We compute the elastic deformation of the sheet
and the corresponding effective gauge field, which modifies the
electronic transport. In a clean system the two-terminal conductance
of the sample is reduced below the ballistic limit and is almost
totally suppressed at low carrier concentrations in samples under
tension. Residual disorder restores a small finite conductivity.

\end{abstract}


\maketitle


{\em Introduction.} Graphene layers which are one or a few carbon atoms
thick~\cite{Netal04,Netal05} combine novel electronic energy spectrum
(``massless Dirac fermions''~\cite{Netal05b,ZTSK05}) and unusual
structural and mechanical
properties~\cite{Metal07,Betal07,Getal08,Betal08,Betal08b} (for a
general review, see Refs.~\onlinecite{GN07,KN07,NGPNG08}). Originally,
graphene sheets lying on quartz substrate have been prepared and
investigated but later it turned out that the freely hanged membranes of
macroscopically large sizes can be derived~\cite{Metal07}. The
electronic transport and mechanical properties of these membranes are
being intensively
studied~\cite{Betal08a,DSBA08,LLA08,Betal07,Getal08,Betal08,Betal08b,
BSHSK08}. They demonstrate, in particular, a much higher electron
mobility, at least, at low temperatures, than graphene sheets on
a substrate~\cite{Betal08a,BSHSK08,DSBA08} and extraordinary mechanical
stiffness~\cite{Betal08}, which makes them especially interesting for
applications. However, peculiarities of the freely hanged
graphene membranes are still poorly understood theoretically.

\begin{figure}
\begin{center}
\includegraphics*[width=3cm]{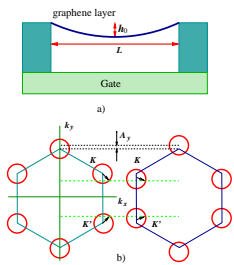}
\end{center}
\caption{(Color online). (a) Sketch of the model of a suspended graphene
sheet under consideration.
(b) Fermi circles positions in the Brillouin zone in the leads (left)
and in the suspended region (right).
\label{sketch}
}
\end{figure}

\begin{figure}
\begin{center}
\includegraphics*[width=4cm]{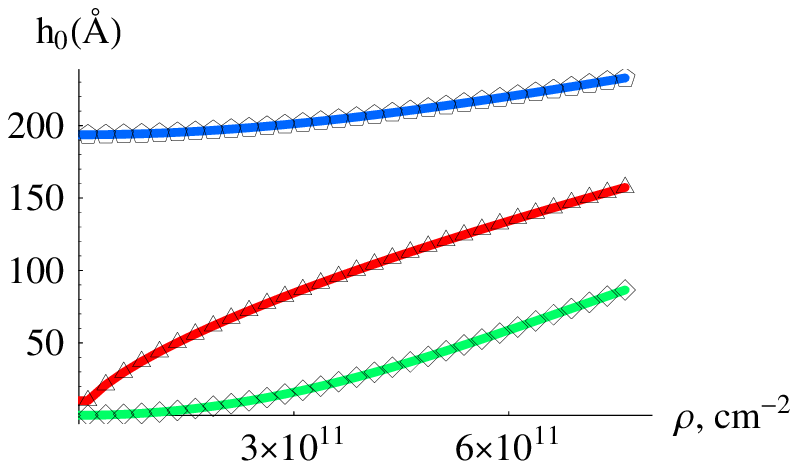}
\includegraphics*[width=4cm]{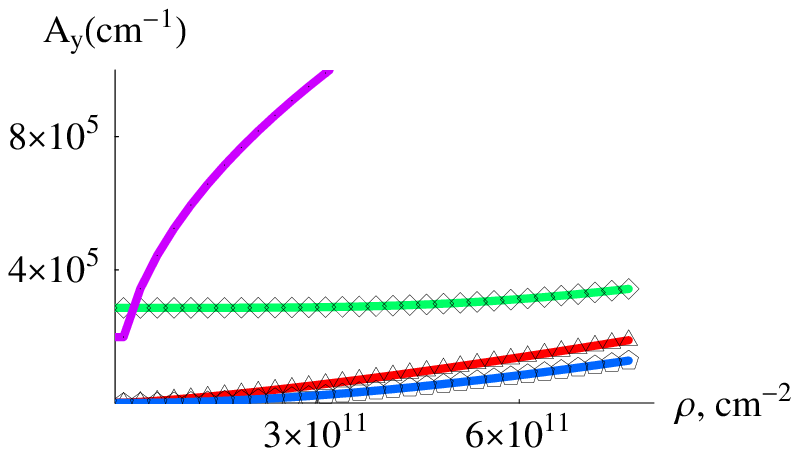}
\end{center}
\caption{(Color online) Left: Maximum height, $h_0$, for a suspended
sheet of length $l = 1 \mu$m as function of carrier density for
different slacks: $\Delta L = 2$nm; pentagons (blue), $\Delta L =
0$, triangles (red), and $\Delta L = -2$nm, diamonds (green). Right:
Effective gauge field for the same three values of slack. The full
magenta curve gives the Fermi wavevector, $k_F$.} \label{height}
\end{figure}

\begin{figure}
\begin{center}
\includegraphics*[width=5cm]{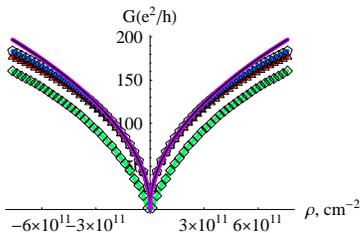}
\end{center}
\caption{(Color online) Conductance as function of carrier density.
The values of $\Delta L$ and the corresponding symbols are the same
as in Fig. 2. The full magenta line gives the ballistic conductance
in the absence of deformation. The width of the sheet is $W =1
\,\mu\text{m}$.} \label{transmission}
\end{figure}

In this paper we demonstrate that unavoidable deformations of the
membranes by an applied electric field can strongly affect their
transport properties. The system we consider is sketched in
Fig.~\ref{sketch}(a). The calculated deformation $h_0$ and
conductance $G$ are shown in Figs.~\ref{height}
and~\ref{transmission}. The left panel of Fig.~\ref{height} shows
the maximum deformation as function of carrier density $n$ and slack
$\Delta L$ in the sheet. We define slack as the difference of the
equilibrium length of the suspended sheet and the distance between
the clamped ends. Negative $\Delta L$ implies that the sheet is
under tension (see below).
 The right panel of Fig.~\ref{height} gives $G$ as function of carrier density for the same three
 values of slack.
In the presence of a finite deformation parametrized by the maximum
vertical displacement $h_0$, the conductance is reduced. We found
that at not too small carrier concentrations $n$, function $G(n)$ is
given by 
\begin{equation}\label{eq:G_fit}
G \simeq \frac{4 e^2}{h} \frac{W}{\pi}
  \left[k_F - \left(\pi - \frac12 \right) |A_y|\, \right]\,,
\end{equation}
where $W$ is the width of the sample, $k_F = \sqrt{\pi |n|}$ is the
Fermi wavevector, and $A_y$ of dimension of inverse length is
related to the deformation [Eq.~(\ref{gauge})]. For graphene
initially under tension~\cite{Betal08b}, where $h_0$ is nearly
constant, and low carrier concentrations, we find a nearly complete
suppression of transport. The physical mechanism of this phenomenon
can be understood as follows. The deformation shifts the Dirac
points by the amount $A_y$ [Fig.~\ref{sketch}(b)], which creates a
mismatch between the graphene leads and the suspended region. If
this shift exceeds the diameter $2 k_F$ of the Fermi circle, the
electrons are fully reflected.

%

Recent experiments~\cite{BSHSK08} show a qualitative agreement with
our Fig.~\ref{transmission}; however, at very low $n$ the
conductivity saturates at at a {\textit finite\,} value $\sim
e^2/h$. We attribute this to residual disorder-assisted tunneling
processes.

{\em The model.} We consider a strip of graphene clamped at two
parallel edges $x = \pm L / 2$, see Fig.~\ref{sketch}. We assume
that the width of the strip in the undeformed state is $L + \Delta
L$, where $\Delta L$ can be of either sign. The strip is suspended
above a control gate, whose electric field induces electron
concentration $n$ in graphene and exerts on it the pressure $P = ( 2
\pi e^2 n^2 ) / \epsilon$.
%
%

The resultant height profile $h(x)$ of the strip has a well known analytic
form~\cite{TW59} under the condition $h_{0} \equiv \max h(x) \ll L$ where the
linear elasticity theory applies.
We focus on the case where $n$ is either comparable or much larger
than $n_0 = \sqrt{(16/\pi) (\epsilon \kappa )/( e^2 L^3 )} \approx
(6 \times 10^9\,\text{cm}^{-2})/(L / 1\,\mu\text{m})^{3/2}$, where
$\kappa \approx 1.1\,\text{eV}$ is the bending rigidity of
graphene~\cite{Fetal07}.
%
%
In this case the deformation is nearly parabolic:
\begin{equation}
\label{profile_1} h(x) \simeq h_{0} \left(1 - \frac{4 x^2}{L^2}
\right)\,, \quad \frac{L}{2} - |x| \gg L\, \frac{\ln u}{u}\,.
\end{equation}
where $u = ({n}/{n_0}) ({L} / {h_{0}})^{1 / 2} \gg 1$.
%
The maximum deformation $h_{0}$ is the positive root of the cubic equation
\begin{equation}
\label{eq:h_0}
\left(h_{0}^2 - \frac{3}{16} L \Delta L \right) h_{0}
= \frac{3 \pi}{64} \frac{e^2}{\epsilon E}\, (n L^2)^2\,,
\end{equation}
where $E \approx 22\,\text{eV}/\text{\AA}^2$ is the Young's modulus
of graphene~\cite{Betal08}. The values of $h_0$ for different values
of the slack are plotted in Fig.~\ref{height}.

The deformation of the strip induces perturbations of two types acting on
electrons: the scalar potential $V(x)$ and the effective vector
potential~\cite{Metal06,M07} $\mathbf{A}(x)$. We examine these potentials below,
starting with $\mathbf{A}(x)$.

{\em Vector potential\/}. The aforementioned shift of the Dirac
points [Fig.~\ref{sketch}(b)] is equivalent to the effect of a
constant vector potential $\textbf{A} = (A_x, A_y)$. We use this
latter formalism in the following as it can be easily generalized to
more complex situations. The role of $\textbf{A}$ is the largest
when the ``zigzag'' direction is along the $y$-axis. Assuming this
is the case, we obtain
\begin{equation}\label{gauge}
A_x(x) = 0\,,\quad A_y(x) = C_1 \xi \frac{\beta}{a} \frac{t}{E} =
C_1 \xi \frac{\beta}{a}\frac{P L^2}{8 E h_0}\,,
\end{equation}
where $\beta = d \log ( \gamma_0 ) / d \log ( a ) \approx 2$ is the
dimensionless electron-phonon coupling parameter, $\gamma_0 \approx
3\, \text{eV}$ is the nearest neighbor hopping, $a \approx
1.4\,\text{\AA}$ is the distance between nearest carbon atoms, $\xi
= \pm 1$ is the valley index, $t \simeq P L^2 / (8 h_0)$ is the
horizontal component of tension per unit length at the edges, and
$C_1$ is a parameter of order unity which depends on the relative
displacements within the unit cell, determined by the microscopic
force constants\cite{SA02b}. (Note that $u = L \sqrt{t / 4
\kappa}\,$.) The corresponding effective magnetic field $B(x) =
-\partial_x A_y$ consists of two narrow spikes at the edges $x = \pm
L / 2$~\cite{GHL08b}.


{\em Transport.} To compute the two-terminal conductance $G$ through a
graphene sheet we assume perfect semi-infinite graphene leads of the
same chemical potential at both ends of the strip. Since the
perturbations depend only on $x$, the $k_y$ momentum is conserved.
However, the effective magnetic field at the edges shifts the mechanical
momentum, $k_y \to k_y - A_y \text{sign}(x)$, of electrons that enter
the strip. This leads to a nearly complete reflection of electrons at
low density where $k_F < A_y / 2$. Similar effect has been previously
examined in the context of transmission of Dirac particles through a
region of homogeneous magnetic field~\cite{Demar07}. However, the
present problem has new qualitative features, see below.

\begin{figure}
\begin{center}
\includegraphics*[width=5cm]{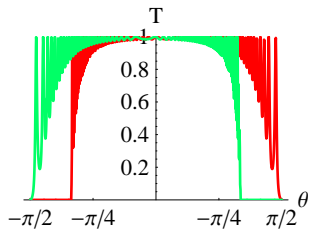}
\end{center}
\caption{(Color online). Angular dependence of the transmission for
$h_{0} = 20\,\text{nm}$, no slack, and $L = 1\,\mu\text{m}$. The
carrier density is $n = 2 \times 10^{11}\,\text{cm}^{-2}$. The two
curves correspond to the two inequivalent Dirac
points.}\label{transmission_angle}
\end{figure}

For a constant $A_y$, the transmission coefficient $T(k_y)$
can be computed analytically:
\begin{equation}\label{T_A}
T(k_y) = \frac{k(0)^2 k(A_y)^2}{k(0)^2 k(A_y)^2 + k_F^2 A_y^2 \sin^2 [k(A_y) L]}\,,
\end{equation}
where $k(q) = \sqrt{k_F^2 - (k_y - q)^2}$. If $k(A_y)^2 < 0$, then $k(A_y)$ is
pure imaginary, so that $\sin^2 [k(A_y) L] \to - \sinh^2 |k(A_y) L|$. In this
case $T(k_y)$ is exponentially small. The plot of $T(k_y)$ is shown in
Fig.~\ref{transmission_angle} using the parametrization $k_y = k_F \sin \theta$
for $-\pi / 2 < \theta < \pi / 2$ and $n = 2 \times 10^{11}\,\text{cm}^{-2}$.
The transmission is indeed almost zero for a range of incident angles $\theta$.
In addition, we see Fabry-P\'erot resonances because of the multiple scattering
off the two interfaces. (Such resonances are essentially absent if the field is
uniform~\cite{Demar07}.)

Neglecting the contribution of edge channels, which is permissible when the
number of bulk channels $k_F W / \pi$ is large, the conductance can be
computed from
\begin{equation}\label{eq:G}
G = \frac{4 e^2}{h} W
    \int\limits_{-k_F}^{k_F} \frac{d k_y}{2 \pi} T(k_y)\,.
\end{equation}
This integral can be done analytically in the limit $k_F \gg |A_y|$
and $L k_F \gg 1$, see  Eq.~\ref{eq:G_fit}. Using Eqs.~(3)--(6), we
also computed $G(n)$ numerically for three $\Delta L$ shown in
Fig.~3. For $\Delta L \geq 0$, where the tension and therefore
$|A_y|$ rapidly increase with $n$ (Fig.~1), the deviations from the
ballistic formula occur at large $n$. In contrast, for $\Delta L
=-2\,\text{nm}$, where the tension is approximately constant, the
largest effect of the gauge field is felt at low concentrations,
leading to a nearly complete vanishing of $G$ at $|n| < 2.0 \times
10^{10}\,\text{cm}^{-2}$.

%
%

{\em Disorder effects.} Any realistic system contains disorder. When weak, it
does not change qualitatively the above results for $G(n)$ at high carrier
concentration $|n| \gg A_y^2$. In the opposite limit, when $G$ is strongly
suppressed, the effect of disorder is important because it relaxes the
constraint of momentum conservation, compensating for the momentum shift due to
the gauge field at the interfaces $x = \pm L / 2$.

Consider the experimentally relevant case~\cite{BSHSK08} where the elastic
mean-free path $l$ is comparable to the system size $L$. Then, at $1
\ll G / (4 e^2 / h) \ll k_F W$ we have a regime where the disorder is weak enough
that its effect on the average conductance is still negligible yet it is strong
enough to fully mix the transverse modes in the suspended region. In this case
the conductance is limited by the two interfaces, which act as classical
resistors in series~\cite{B97}, $G = G_i / 2$. Let us compute the
conductance of a single interface $G_i$ with and without including the effect of
disorder.

If disorder \textit{near the interface} is neglected, $G_i$ is given by the
expression similar to Eq.~(\ref{eq:G}) where $T(k_y)$ is now the transmission
coefficient through a single interface:
$T(k_y) = [4 k(0) k(A)]/[ (k(0) + k(A))^2 + A_y^2]\,$.
This formula holds if $k_y$-mode is propagating, $\Im\text{m} k(0) =
\Im\text{m} k(A_y) = 0$. Otherwise, it is evanescent and $T(k_y) = 0$. Each
evanescent mode decays exponentially either to the left or to the right of $x =
0$ interface, depending on which of $\Im\text{m} k(0)$ and $\Im\text{m} k(A_y)$ is
nonzero. 
At $k_F < |A_y| / 2$ all modes at the Fermi energy are evanescent. Therefore,
if the disorder is neglected, $G_i$
vanishes.

Let us now include disorder-induced mixing among the evanescent and propagating
modes, which gives a correction $\Delta T(k_y)$ to each $T(k_y)$. To the lowest
order in the concentration $n_{s}$ of scatterers, this correction can be written as
\begin{equation}\label{eq:DeltaT}
\Delta T(k_y) = \int\limits_{A_y - k_F}^{A_y + k_F}
                  \frac{d k_y^\prime}{2 \pi}\,
                  \int\limits_{-\infty}^{\infty} d x n_{s}
                  \delta T(k_y, k_y^\prime, x)\,,
\end{equation}
where $\delta T(k_y, k_y^\prime, x)$ is the off-diagonal transmission
coefficient due to a single scatterer at position $\mathbf{r} = (x, y)$. Function
$\Delta T(k_y)$ has the dimension of length, similar to the transport
cross-section $\Lambda_{s} = 1 / (n_{s} l)$. On physical grounds, we expect
$\delta T(k_y, k_y^\prime, x) \sim (\Lambda_{s} / k_F) \exp\left(-2
|k(A_y) x| \right)$.
Here the exponential represents the probability of the evanescent wave
to reach the scatterer and the prefactor provides the correct units and
scaling with disorder strength. The dominant contribution to $\Delta
T(k_y)$ comes from the scatterers located in the strip $|x| \lesssim 1 /
|k(A_y)|$. Integrating over $x$, $k_y$, and $k_y^\prime$, we finally
get
\begin{equation}\label{eq:G_i_asym}
G = \frac{G_i}{2} = \frac{4 e^2}{h}\, \frac{k_F W}{\pi}
                    \frac{C_2}{2 |A_y| l}
 = \left(\frac{e^2}{h} \right)^2 \frac{4 C_2 W n }{|A_y| \sigma(n)}\,,
\end{equation}
where $C_2$ is a numerical coefficient and $\sigma(n)$ is the
conductivity of a sample with size $L \gg l$. A formal derivation based
on the Green's function formalism yields $C_2 = 4$. For estimate,
we can take $n = 10^9\,\text{cm}^{-2}$, $\sigma \sim 4 e^2 /
h$, $W = 1\,\mu\text{m}$, and $A_y = 2 \times 10^5\,\text{cm}^{-1}$. We
then find $G \sim 2 e^2 / h$, i.e., an appreciably large value.

{\em Scalar potential\/}. The deformation of the graphene strip also creates a
scalar potential $V(x)$ in the system. Our estimates below indicate that it
is relatively small, so that its presence is not expected to change the results
shown in Figs.~\ref{transmission} and \ref{transmission_angle} in a major way.
We will discuss it briefly, for completeness.

The bare potential induced by the deformation is:
\begin{equation} \label{eq:scalar}
\begin{split}
V_\text{ext}(x) &= -\frac{P h ( x )}{e n}
      + V_0  \left(u_{xx} + u_{yy} \right)\\
 &= -\frac{P h ( x )}{e n} + \theta
\left(\frac{L}{2} - |x| \right) \frac{t V_0}{E}\,,
\end{split}
\end{equation}
where the first term is the change in electrostatic potential due to
the change in distance to the gate, the second term gives the
deformation potential induced by a local
compression~\cite{OS66,SA02b} $u_{xx} + u_{yy}$, and $V_0 \approx
10\,\text{eV}$.

Within the linear screening theory, the Fourier transform $\tilde{V}$ of
$V$ is given by $\tilde{V}(q) = \tilde{V}_{ext} ( q ) / \varepsilon
(q)$, where $\varepsilon(q)$ is the dielectric function. For reasonable
carrier concentrations, the potential in Eq.~(\ref{eq:scalar}) is smooth
over distances $\sim k_F^{-1} \ll L$. We can use the Thomas-Fermi (TF)
approximation, $\varepsilon(q) = 1 + k_s /\, |q|$, where $k_s = 4 \alpha
k_F$ is the inverse TF screening length and $\alpha = e^2 / \epsilon
\hbar v \sim 1$ is the dimensionless strength of the Coulomb
interaction. The screened potential can be computed analytically in
terms of special functions. In the limit $k_s L \gg 1$ and at distances
greater than $k_s^{-1}$ from the boundaries it reads:
\begin{equation} \label{eq:V_e}
\begin{split}
 e V (x) &\simeq \frac{1}{2 \pi}\,
                 \frac{P V_0}{E h_0 k_s}
                 \frac{L^3}{L^2 - 4 x^2}\\
&- \frac{8}{\pi} \frac{P h_0}{n k_s L} \left(
1 + \frac{x}{L} \ln \left| \frac{L - 2 x}{L + 2 x} \right| \right)\,.
\end{split}
\end{equation}
The potential at the edge is given by
\begin{equation}
\label{eq:V_edge} e {V} \left( \frac{L}{2} \pm 0\right)
 =  \pm \frac{P V_0 L^2}{16 E h_0}
 + \frac{4}{\pi} \frac{P h_0}{n k_s L}
   \left[C_3 - \ln (k_s L) \right]\,,
\end{equation}
where $C_3 \sim 1$. Thus, the divergences in Eq.~(\ref{eq:V_e}) are cut
off at the distance of the order of the screening length $1 / k_s$ from
the edge, as expected.

For $n \sim 3 \times 10^{11}\,\text{cm}^{-2}$, the deformation and
the electrostatic potentials at the edges are comparable in
magnitude and together amount to about $10\%$ of the Fermi energy.
Although this is not a negligible amount, it is still numerically
small, so that the linear-response screening approach is justified.


{\em Experimental implications and future directions.} In this paper
we focused on a simplest geometry of a suspended graphene strip
(Fig.~\ref{sketch}) for which the calculation of the transport
properties can be done semi-analytically. We have found that
reasonable values of the tension, $\Delta L  / L \sim 0.2 \%$ in a
suspended graphene strip can lead to the almost total reflection for
a significant range of incoming momenta, which causes a downward
shift of the two-terminal conductance compared to the ballistic
limit, Eq.~(\ref{eq:G_fit}). Residual disorder can partially
compensate for this shift, which may be the case in current
experiments~\cite{Betal08a,BSHSK08,DSBA08}. Note however, that our
theory cannot be directly compared with experiments, as the sample
geometry and therefore the configuration of the gauge field can be
more complicated than what we have assumed here.

There are a number of possible directions for future study. One
interesting problem is how the deformation would affect the quantum Hall
effect (QHE) in the suspended graphene. Below we offer a
preliminary discussion of this question.

For the model considered, the effects of the deformation are
restricted to the $x = \pm L / 2$ interfaces. The Landau levels near
such lines will be modified (except for the $N = 0$ one, which is
topologically protected~\cite{Gies07,GKV08}). Quasiclassicaly, the
reflection at the interfaces creates skipping orbits, which
propagate parallel to the $y$-axis but in opposite directions on the
two sides of each interface. This could lead to backscattering of
the edge currents and modification of the QHE. The effect is the
strongest when $k_F \lesssim |A_y|$. For low-lying Landau levels,
where $k_F \sim 1 / l_B$, an estimate of the external magnetic field
$B_*$ below which the QHE is affected can be derived from the
condition $l_B(B_*) \sim |A_y|^{-1}$. For $\Delta L = - 2$nm, and
$A_y \sim 2 \times 10^5$ cm$^{-1}$, it yields $B_* \sim
0.7\,\text{T}$.

In a more realistic geometry, a small fictitious magnetic field will
also exist inside the suspended region. Its magnitude  is of the
order of $0.05\,\text{T}$ for the same $\Delta L$ and $L$. Landau
levels mixing in the bulk occurs when the external field is
comparable or smaller than this value.




This work was supported by MEC (Spain) through grant
FIS2005-05478-C02-01 and CONSOLIDER CSD2007-00010, the Comunidad de
Madrid, through CITECNOMIK, CM2006-S-0505-ESP-0337, the EU Contract
12881 (NEST), the Stichting voor Fundamenteel Onderzoek der Materie
(FOM) (the Netherlands), and by the US NSF under grant DMR-0706654. We
appreciate helpful conversations with M. A. H. Vozmediano.

\bibliography{bib_suspended_3}
\end{document}